\documentclass[review]{elsarticle}
\usepackage{lineno,hyperref}
\usepackage[normalem]{ulem}
\usepackage{graphics}
\usepackage{amsmath,bm}
\usepackage{amssymb}
\usepackage{graphicx}
\usepackage{bm}
\usepackage{amsfonts}
\usepackage{color}
\usepackage{epstopdf}
\usepackage{booktabs}
\usepackage{array}

\usepackage{lineno,hyperref}
\usepackage{graphicx}
\usepackage{hyperref}
\usepackage{graphics}
\usepackage{epsfig}
\usepackage{bm}
\usepackage{epic,eepic}
\usepackage{epsfig}
\usepackage{pifont}

\usepackage{subfigure}

\usepackage{xspace}
\usepackage{graphicx,color}
\usepackage{amsmath,amssymb,amsfonts,mathrsfs}
\usepackage{url}
\usepackage[normalem]{ulem}
\usepackage{booktabs}
\usepackage{float}

\usepackage[toc,page]{appendix}
\usepackage[capitalize]{cleveref}
\usepackage{mathtools}
\usepackage[normalem]{ulem}
\usepackage{gensymb}

\providecommand{\st}[1]{{}^{_{\text{#1}}}\!}

\newcommand{\be}{\begin{equation}}
\newcommand{\ee}{\end{equation}}
\newcommand{\ben}{\begin{equation*}}
\newcommand{\een}{\end{equation*}}

\newcommand{\bv}[1]{\mathbf{#1}}

\newcommand{\vv}{\bv{u}}

\newcommand{\delF}[1]{\textcolor{blue}{}}

\newcommand{\tdiff}{t_{\mbox{\tiny diff}}}

\newcommand{\rp}{{\mathrm{R_p}}}
\newcommand{\rhop}{\rho_{\mathrm{p}}}

\newcommand{\uconf}{U^{\mathrm{(z)}}_{\mathrm{conf}}}
\newcommand{\uo}{U^{\mathrm{(z)}}_{\mathrm{unconf}}}

\newcommand{\kbt}{{\mathrm{k_B T}}}
\newcommand*{\fg}{\mathrm{g}}
\newcommand*{\fp}{\mathrm{F_p}}
\newcommand*{\Do}{D_{\mathrm{unconf}}}
\newcommand*{\Dconf}{D_{\mathrm{conf}}}
\newcommand*{\gconf}{\gamma_{\mathrm{conf}}}
\newcommand*{\go}{\gamma_{\mathrm{unconf}}}
\newcommand*{\cm}{c_m}

\modulolinenumbers[5]

\journal{Journal of computational science}

\begin{document}

\begin{frontmatter}

\title{A lattice Boltzmann study on Brownian diffusion and friction of a particle in a confined multicomponent fluid}

\author[address1,address2]{Xiao Xue\corref{mycorrespondingauthor}}
\cortext[mycorrespondingauthor]{Corresponding author}
\ead{X.Xue@tue.nl}

\author[address1]{Luca Biferale}
\ead{biferale@roma2.infn.it}

\author[address1]{Mauro Sbragaglia}
\ead{sbragaglia@roma2.infn.it}

\author[address3,address4]{Federico Toschi}
\ead{f.toschi@tue.nl}

\address[address1]{Department of Physics \& INFN, University of Rome ``Tor Vergata'', Via della Ricerca Scientifica 1, 00133, Rome, Italy.}
\address[address2]{Department of Physics  and J.M. Burgerscentrum, Eindhoven University of Technology, 5600 MB Eindhoven, the Netherlands.}
\address[address3]{Departments of Physics and of Mathematics and Computer Science and J.M. Burgerscentrum, Eindhoven University of Technology, 5600 MB Eindhoven, the Netherlands.}
\address[address4]{Istituto per le Applicazioni del Calcolo CNR, Via dei Taurini 19, 00185 Rome, Italy.}

\begin{abstract}
We study the diffusivity of a small particle immersed in a square box filled with a non-ideal multicomponent fluid in the presence of thermal fluctuations. Our approach is based on the numerical integration of fluctuating lattice Boltzmann models (LBM) for multicomponent mixtures. At changing the wettability on the particle's surface, we measure the mean square displacement (MSD) and compare with the prediction of the Stokes-Einstein theory. Two main set-ups are tested, involving periodic boundary conditions and wall boundary conditions realized on the computational box. We find that full periodic boundary conditions give rise to random advection after millions of lattice Boltzmann time steps, while this effect is mitigated in the presence of wall boundary conditions. The matching with the Stokes-Einstein relation is therefore guaranteed when we use the appropriate frictional properties measured in the presence of confinement. Our results will help the exploration of nanoscale applications with multicomponent fluids using LBM in the presence of thermal fluctuations.
\end{abstract}

\begin{keyword}
\texttt{Brownian Diffusion; finite-size particle; fluctuating lattice Boltzmann methods; multicomponent fluids}
\end{keyword}

\end{frontmatter}

%
\section{INTRODUCTION}
The lattice Boltzmann models (LBM), known to bridge the macroscopic~\cite{toschi2009lagrangian} and microscopic scale problems~\cite{ladd1993short}, have been well developed in past few decades~\cite{succi2001lattice,kruger2017lattice}. They have been successfully formulated to study complex fluid phenomena, including turbulence~\cite{teixeira1998incorporating,chen1998lattice,karlin1999perfect},  non-ideal fluids with phase transition and/or segregation~\cite{he1999lattice, liu2012three, reis2007lattice, chiappini2019, chiappini2018ligament, milan2018lattice}, polymer flows~\cite{ahlrichs1999simulation,ahlrichs1998lattice,berk2005lattice}, active matter~\cite{de2016lattice}. Through the Chapman-Enskog expansion, the LBM can recover the hydrodynamic representation of the Navier-Stokes equations~\cite{lamb1993hydrodynamics,milne1996theoretical}. Recent studies have shown interests in integrating finite-size particles into the LBM framework, such as particle suspensions~\cite{ladd1994numerical, aidun1998direct, ladd2001lattice, wu2010simulating, nguyen2002lubrication}, pickering emulsions~\cite{jansen2011bijels}, self assembly particles~\cite{davies2014assembling,xie2015tunable,xie2016controlled}. However, there are relatively few LBM studies in the literature on particle-fluid interaction problems at the nanoscale, where the effects of the thermal fluctuations cannot be neglected. The deterministic hydrodynamics of Navier-Stokes equations is inadequate to describe the flow evolution at the nanoscales~\cite{Landau, Zarate2006}.  For this reason, there have been pioneering works to extend the LBM in the presence of thermal fluctuations, designing the so-called fluctuating lattice Boltzmann models (FLBM)~\cite{Varnik11,Ladd,Adhikari2005,Dunweg07,Gross10,KaehlerWagner13}. Recently, FLBM have been used to study non-ideal multicomponent fluids~\cite{Belardinelli15,Belardinelli19} and also the effects of thermally excited capillary waves on the break-up properties of a thin liquid ligament~\cite{xue2018effects}.  Moreover, FLBM have been used to understand the influence of thermal fluctuations on the particle settling under confinement~\cite{xue2019}. When the particle is coupled with the FLBM, it will naturally perform Brownian motion. In such conditions, the mean squared displacement (MSD) of the particle in an unconfined domain is expected to follow the Einstein's relation in time $t$~\cite{kubo1966fluctuation, risken1996fokker}
\begin{equation}
\left \langle (\mathbf{x}(t) - \mathbf{x}(t_0))^2 \right \rangle = 6\Do t,
\end{equation}
where the particle is located at ${\bf x}$ and the diffusion coefficient in unconfined domains is indicated with $\Do$. Based on the fluctuation-dissipation relation~\cite{kubo1966fluctuation}, the diffusion coefficient $\Do$ can be represented by $\kbt/(6\pi\eta\rp)$, where $\kbt$ is the system thermal energy, $\eta$ is the dynamic viscosity and $\rp$ is the particle radius. Notice that the denominator in $\Do$ is essentially the friction coefficient experienced by a spherical particle with radius $\rp$ in unconfined domains. Due to the large computational cost associated with the study of Brownian motion, only short-time Brownian motion studies of colloids have been proposed in the FLBM framework~\cite{ladd1993short}. The long-time Brownian motion has rarely been studied. In this paper, we aim to quantitatively tackle the problem of the Brownian motion of a wetted particle immersed in a multicomponent fluid. In~\cref{sec:methodology}, we introduce our multicomponent FLBM algorithm which couples with finite-size wetted particles. In~\cref{sec:num-setup}, we present our simulation setup. Results are discussed in~\cref{sec:results}, where we show that the particle experiences a random advection by using periodic boundary condition which does not allow a fair assessment of the Einstein's relation with the diffusion coefficient predicted by the fluctuation-dissipation relation~\cite{kubo1966fluctuation}. We show how to remove this pathology when working with wall boundary conditions; the precise matching with the Einstein's relation, however, requires the knowledge of the friction in the presence of confinement. We {draw our} conclusions in~\cref{sec:conclusions}
\section{METHODOLOGY}\label{sec:methodology}
In this section, we introduce the multicomponent FLBM for the binary mixture of two fluids~\cite{Belardinelli15}. Also, the wetted finite-size particle model~\cite{ladd1994numerical,aidun1998direct,jansen2011bijels}, which is used to simulate the solid particle, is coupled with the fluctuating multicomponent fluids. Technical details of the FLBM have already been extensively presented in~\cite{Belardinelli15}, and here we only briefly recall the most important aspects for the sake of completeness. We consider $f_{l i}(\mathbf{x},t)$ as the discretized particle's probability distribution function on the $i$-th direction of a lattice cell with velocity $\bv{c}_i$. $l$ represents the fluid component $l=A,B$, and the lattice cell is located at $\mathbf{x}$ at time $t$. In this paper, we employ the D3Q19 model, {corresponding to a} 3D LBM model with discretized 19 velocity directions $\bv{c}_i$ ($i=0...Q-1$). The LBM evolution equation for the binary fluid, which considers collision, fluid-fluid interactions, and stochastic noise, can be written as:\\
\be
\label{eq:lbe}
f_{l i}(\mathbf{x}+\mathbf{c}_{i},t+1) -f_{l i}(\mathbf{x}, t)=\Omega\left[f_{l i}(\mathbf{x},t )-f_{li}^{\mbox{\tiny eq}}(\mathbf{x},t )\right] + F_{l i}(\mathbf{x},t) + \xi _{l i}(\mathbf{x},t),  \hspace{.2in} l=A,B,\\
\ee
where $\Omega$ is a collision kernel~\cite{succi2001lattice, kruger2017lattice}, $F_{l i}$ is forcing term for fluid-fluid interactions, and $\xi _{li}$ is a stochastic source. The collision kernel is chosen to be the multiple relaxation time (MRT) collision kernel~\cite{DHumieres02,Dunweg07,SchillerThesis}. The basic idea of the MRT kernel is to introduce a moment space to decompose the probability functions into ``modes". The lower-order modes are related to hydrodynamic quantities (density, momentum and stress tensor). {The higher-order modes are the ``ghost modes"}, which do not contribute to the hydrodynamic behavior~\cite{DHumieres02,Dunweg07}. The collision kernel relaxes the distribution function towards the local equilibrium $f_{li}^{\mbox{\tiny eq}}$, which is given by
\be
\label{eq:local_eq}
f_{li}^{\mbox{\tiny eq}}\left(\textbf{x},t\right) =  \omega_{i}\rho_l\left(\textbf{x},t\right) \bigg[1+\frac{\textbf{c}_i\cdot\textbf{u}\left(\textbf{x},t\right)}{c_s^2}+\frac{\left[\textbf{c}_i\cdot\textbf{u}\left(\textbf{x},t\right)\right]^2}{2c_s^4}  - \frac{\left[\textbf{u}\left(\textbf{x},t\right)\cdot\textbf{u}\left(\textbf{x},t\right)\right]}{2c_s^2}\bigg] \; ,
\ee
in which $\omega_{i}$ is a suitable weight needed to impose the isotropy in the interaction, $\rho_l\left(\textbf{x},t\right)$ and $\textbf{u}\left(\textbf{x},t\right)$ are the hydrodynamic macroscopic density for each component and mixture velocity, respectively, which can be calculated from the distribution function:
\begin{equation}\label{eq:density}
\rho_{l}(\mathbf{x}, t) = \sum_{i=0}^{Q-1} f_{l i}(\mathbf{x}, t),  \hspace{.2in} \vv(\mathbf{x}, t) = \frac{\sum_{i=0}^{Q-1}\sum_{l} f_{l i}(\mathbf{x}, t)\mathbf{c}_{i}}{\rho_{\st{tot}}(\mathbf{x}, t)},
\end{equation}
where $\rho_{\st{tot}}(\mathbf{x},t)=\sum_l \rho_{l}(\mathbf{x},t)$ is the total density of the two components. 
The term $F_{l i}$ is the Shan-Chen forcing term which models the non-ideal interactions for the mixture~\cite{SC93,SC94,Zhang11,SbragagliaBelardinelli,SegaSbragaglia13}. The Shan-Chen forcing term is given by
\begin{equation}\label{sc-force_eq}
F_{l}(\mathbf{x},t) = - {\cal G}\varphi_{l}(\mathbf{x},t) \sum_{l' \neq l} \sum_{i=0}^{Q-1}\omega _{i}\varphi _{l'}(\mathbf{x}+\mathbf{c}_{i},t) \mathbf{c}_{i}  \hspace{.2in}
\end{equation}
where ${\cal G}$ is a strength coefficient, $\varphi_l$ is known as the pseudo-potential function which is set to the fluid density $\varphi_l = \rho_l$. In all the simulations performed, the coupling coefficient is set to ${\cal G}= 1.5$ lattice Boltzmann units (lbu). The term $\xi_{l i}(\mathbf{x},t)$ in~\cref{eq:lbe} is a stochastic force. The noise term does not influence on the conserved density modes, while higher modes receive the stochastic source following the fluctuation-dissipation relation~\cite{Belardinelli15}. Through the Chapman-Enskog expansion~\cite{Dunweg07,SchillerThesis}, we can recover the equations for the fluid densities and the hydrodynamical velocity~\cite{Zarate2006} (superscript $T$ means transposition):
\be
\partial_t \rho_{\st{tot}} + {\bf \nabla} \cdot (\rho_{\st{tot}} {\vv}^{(H)})=0,
\ee
\be
\partial_t \rho_A+{\bf \nabla} \cdot (\rho_A \vv^{(H)})={\bf \nabla} \cdot \left[ {M} {\bf \nabla} \mu + \bm{\Psi}_{\st{v}} \right],
\ee
\be
\partial_t (\rho_{\st{tot}} {\vv}^{(H)})+{\bf \nabla} (\rho_{\st{tot}} {\vv}^{(H)} {\vv}^{(H)}) = -{\bf \nabla} P_b + {\bf \nabla} \cdot [\eta ({\bf \nabla} {\vv}^{(H)} + ({\bf \nabla} {\vv}^{(H)})^T )+{\bf \Sigma}_{\st{t}}],
\ee
where the hydrodynamical velocity ${\vv}^{(H)}$ is given by ${\vv}^{(H)}={\vv}+({\bf F}_{A}+{\bf F}_{B})/2 \rho_{\st{tot}}$, $P_b$ and $\mu$ are the bulk pressure and the chemical potential, respectively~\cite{SegaSbragaglia13}. The mass diffusion coefficient ${M}$ and $\eta$ are related to the relaxation times of the fluid. In all the simulations performed, the mass diffusion coefficient is set to ${M}=1/6$ lbu, and the dynamic viscosity is set to $\eta=0.383$ lbu. {We use the same kinematic viscosity in both fluids (i.e. the same relaxation time in the LBM framework)}. The stochastic stress (${\bf \Sigma}_{\st{t}}$) and the stochastic diffusion ($\bm{\Psi}_{\st{v}}$) contribution to the equations of hydrodynamics are given by
\be
{\bf \Sigma}_{\st{t}}=\sqrt{\eta \kbt}({\bf W_{\st{t}}}+{\bf W}_{\st{t}}^T) \hspace{.2in} \bm{\Psi}_{\st{v}}=\sqrt{2 D  \kbt} {\bf W_{\st{v}}}
\ee
where $\kbt$ is the thermal energy, while and ${\bf W}_{\st{t}}$ and ${\bf W}_{\st{v}}$ are a Gaussian tensor and a Gaussian vector with independent and uncorrelated components and variance equal to unity. 
\begin{figure}[h]
\centering
\includegraphics[width=0.7\textwidth]{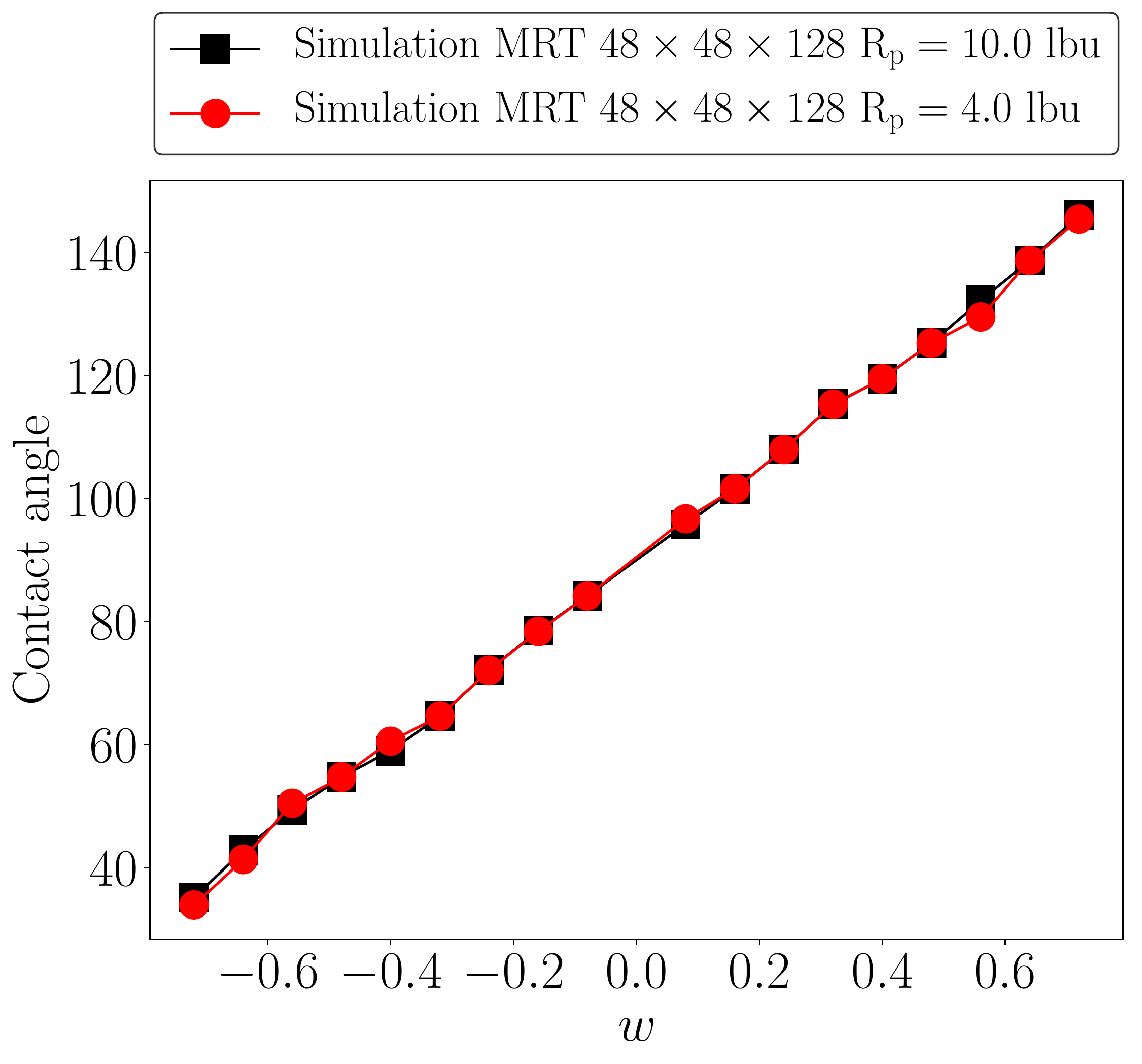}
\caption{Contact angle as function of the tunable parameter $w$ for the wettability of the particle. When $w=0$, a neutral wetting condition is modelled with contact angle equal to $90^{\circ}$. When $w>0$ (hydrophobic case), the particle's contact angle is larger than $90^{\circ}$; when $w<0$ (hydrophilic case), the particle's contact angle is smaller than $90^{\circ}$. Black and red colors refer to two different resolutions for the particle radius ($\rp=10,4$ lbu). {Small particle radii are enough to retrieve converged results.}}
\label{fig:wetting_test}
\end{figure}
For the particle model, the technical details have been introduced in {previous} studies, readers can follow references~\cite{ladd1994numerical,aidun1998direct,jansen2011bijels}. {Here we recall the basic ingredients}. We integrate the wetted finite-size particle in the FLBM framework. The particle is built on lattices by declaring nodes belonging to the particle (``particle nodes''). The particle exchanges the momentum with surrounding fluids through the bounce back~\cite{ladd1994numerical}. {The particle moves due to the action of the external forces on it; due to this motion, new lattice nodes that were originally belonging to the fluid will belong to the particle (cover-node behavior). Consistently, new fluid nodes are generated which were initially belonging to the particle (uncover-node behavior)}. To impose the total mass conservation, the mass correction algorithm described in Ref.~\cite{jansen2011bijels} has been implemented in this paper. We introduce the virtual fluid in the outer-most layer of the particle for tuning the particle's wettability. The {densities in the} virtual fluid nodes for the two components are defined as $\rho_{A, v}$ and $\rho_{B, v}$. The parameter $w$ is a dimensionless number that regulates the affinity of the particle towards one of the two fluids~\cite{jansen2011bijels}. {The wettability properties described in the following (i.e. hydrophobic, neutral, hydrophilic) refer to the affinity of the particle towards the majority component in the bulk phase. A positive (negative) $w$ will correspond to a hydrophobic (hydrophilic) case. We also placed} the wetted finite-size particle at the fluid-fluid interface and measured the contact angle of the particle by tuning the parameter $w$. ~\cref{fig:wetting_test} shows the contact angle as function of $w$. Two different radii of {the} particle have been tested: $\rp = 4$ lbu (red line) and $\rp = 10$lbu (blue line). The results show that the particle {wettability} can be tuned from around $30^{\circ}$ contact angle to $150^{\circ}$ contact angle.
\section{NUMERICAL SETUP }\label{sec:num-setup}
%
\begin{figure}[h]
\centering
\includegraphics[width=1\textwidth]{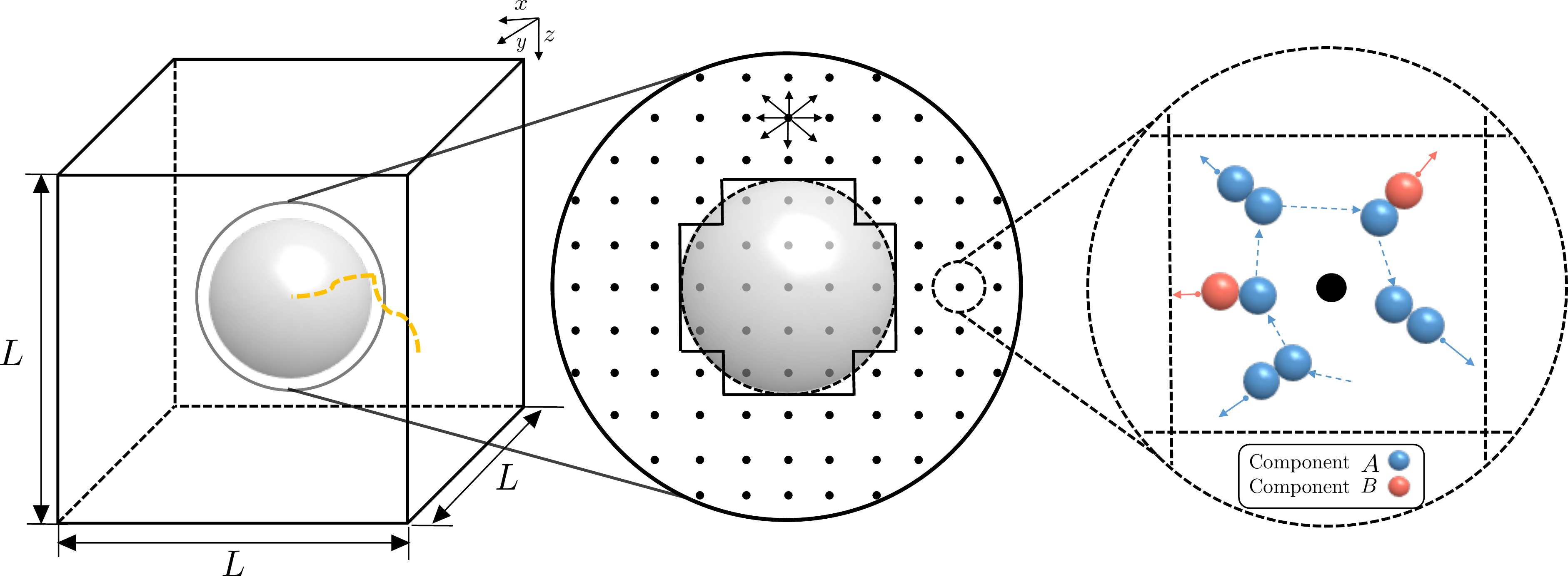}
\caption{Sketch of the setup for the Brownian motion of the wetted particle in the cubic box. The box is an isometric domain with a size of $L \times L \times L$. The solvent is a fluctuating mixture of the two components, A and B, with the majority of component A in the bulk phase. The solvent solver is based on the implementation of a multicomponent {fluctuating lattice Boltzmann model} (FLBM)~\cite{Belardinelli15}. The wetted particle is coupled with the binary fluid~\cite{aidun1998direct, jansen2011bijels}.} \label{fig:sketch_brownian}
\end{figure}
\begin{figure}[h!]
\centering
\includegraphics[width=1\textwidth]{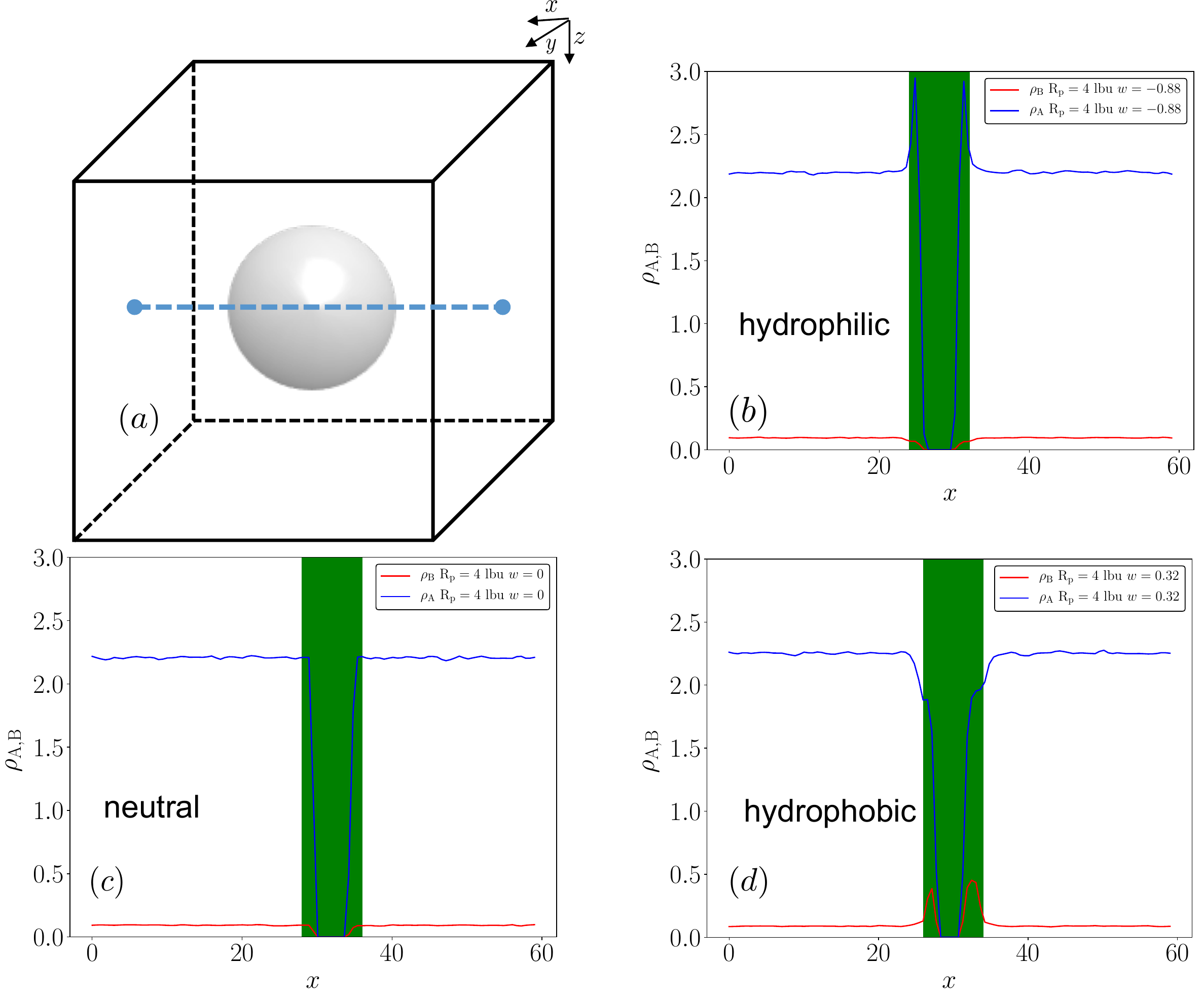}
\caption{{We report the density profiles of the two components along an axial line (Panel (a)) at changing the wettability boundary conditions: hydrophilic (Panel (b)); neutral (Panel (c)) and hydrophobic (Panel (d)). The wettability property refers to the majority component in the bulk phase (component B).}}\label{fig:particle_contactline}
\end{figure}

The numerical set-up for the Brownian motion simulation is shown in~\cref{fig:sketch_brownian}. We place a particle with radius $\rp$ in the cubic box with domain size $L \times L \times L$, where $L=60$ lbu. The domain boundary conditions can be periodic boundaries or neutral wetting walls. The box is filled with the two species of fluids ($A$ and $B$) {in a mixed state}. The component $A$ is the majority {component of the mixture} with density $\rho_{\mathrm{A}} = 2.21$ lbu in the bulk, while component $B$ is the minority component with density $\rho_{\mathrm{B}} = 0.09$ lbu in the bulk. {As described previously,} the particle's {wettability} can be tuned as hydrophilic ({increased affinity for} component $A$), neutral (no preference) or hydrophobic ({increased affinity for} component $B$). {\cref{fig:particle_contactline} shows the concentration of the two components in proximity of the particle's surface and in the bulk. The density of the particle is set to $\rhop = 2\rho_{\st{tot}}$. We define the intrinsic diffusion timescale $\tdiff$ as the time it takes for fluid perturbations to diffuse over a lengthscale comparable to a particle radius~\cite{ladd1993short}}
\begin{equation}
{\tdiff = \rp^2/\nu}.
\end{equation}
\begin{figure}[t!]
\centering
\includegraphics[width=0.7\textwidth]{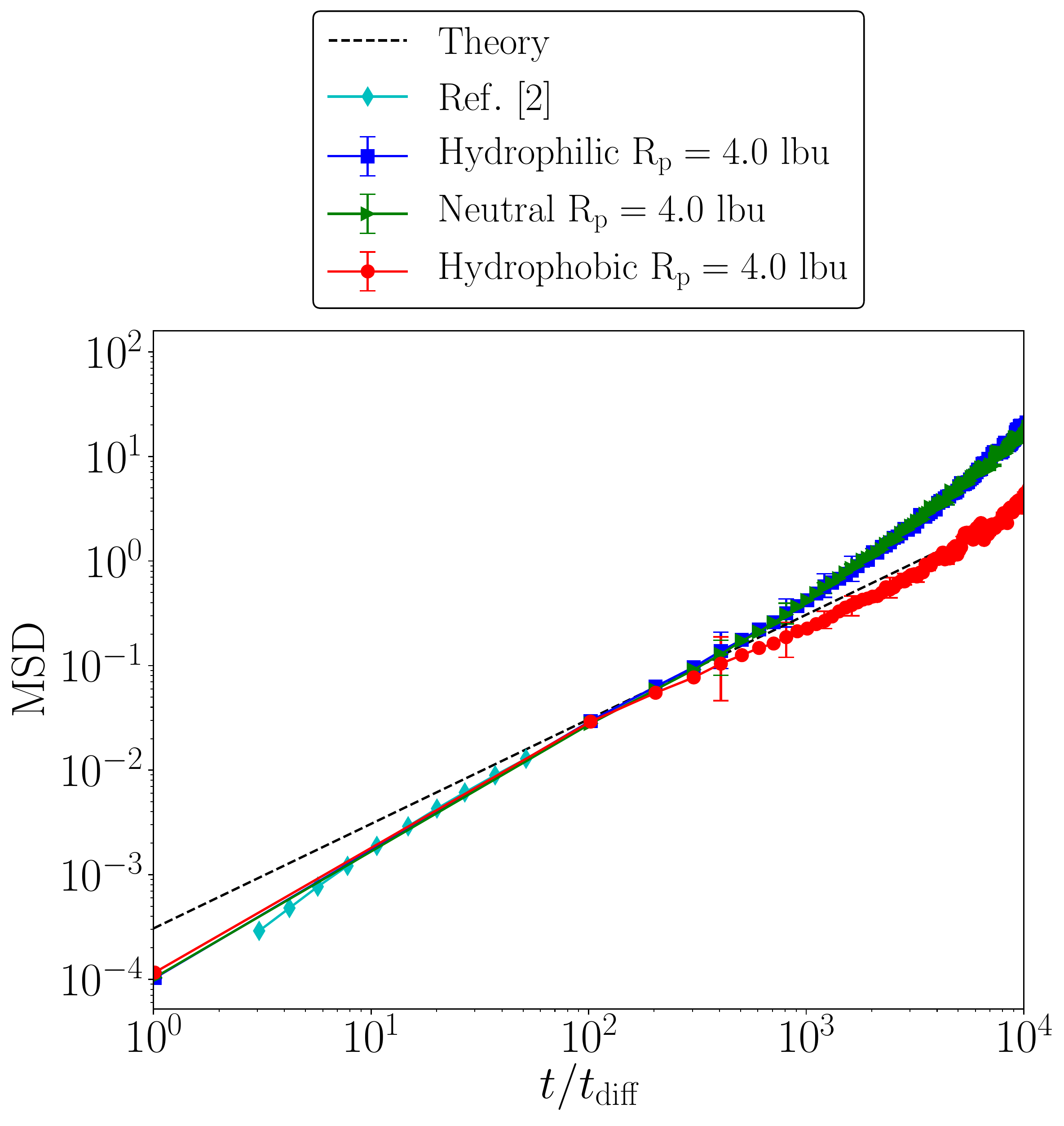}
\caption{Mean square displacement $MSD$ as a function of the dimensionless time unit $t/\tdiff$ at changing the wetting parameter $w$. Blue, green, and red colors represent wetted particle equipped with hydrophilic($w=-0.88$), neutral($w=0$) and hydrophobic cases($w=0.32$). Cyan color represent the short-time Brownian motion from Ref.~\cite{ladd1993short}. The particle's MSD does not match the Einstein's prediction when $t/\tdiff > 100$. Errorbars are the standard deviation estimated from different Brownian motion experiments.}\label{fig:particle_msd_periodic}
\end{figure}
To perform simulations and match the asymptotic Brownian motion behavior with the Einstein's prediction, we need at least $10^4$ time units. Thus, we cannot use very high resolution for the particle and we chose the radius of the particle equal to $\rp= 4$  lbu. The domain size is set to $60^3$ lbu, and the timescale $\tdiff$ is $\tdiff=96$ lbu. The wetting parameters of the particle are set to $w=-0.88$ (hydrophilic), $w=0$ (neutral), and $w=0.32$ (hydrophobic). The thermal energy is set to $\kbt = 1\cdot 10 ^{-4}$ lbu. For each wetting parameter, we perform simulations with $10^7$ lbu timesteps with {10 realizations of the noise} to get enough statistics. 
\section{RESULTS}\label{sec:results}
\begin{figure}[t]
\begin{minipage}{1.0\textwidth}
\centering
\subfigure{\includegraphics[width = 0.7\linewidth] {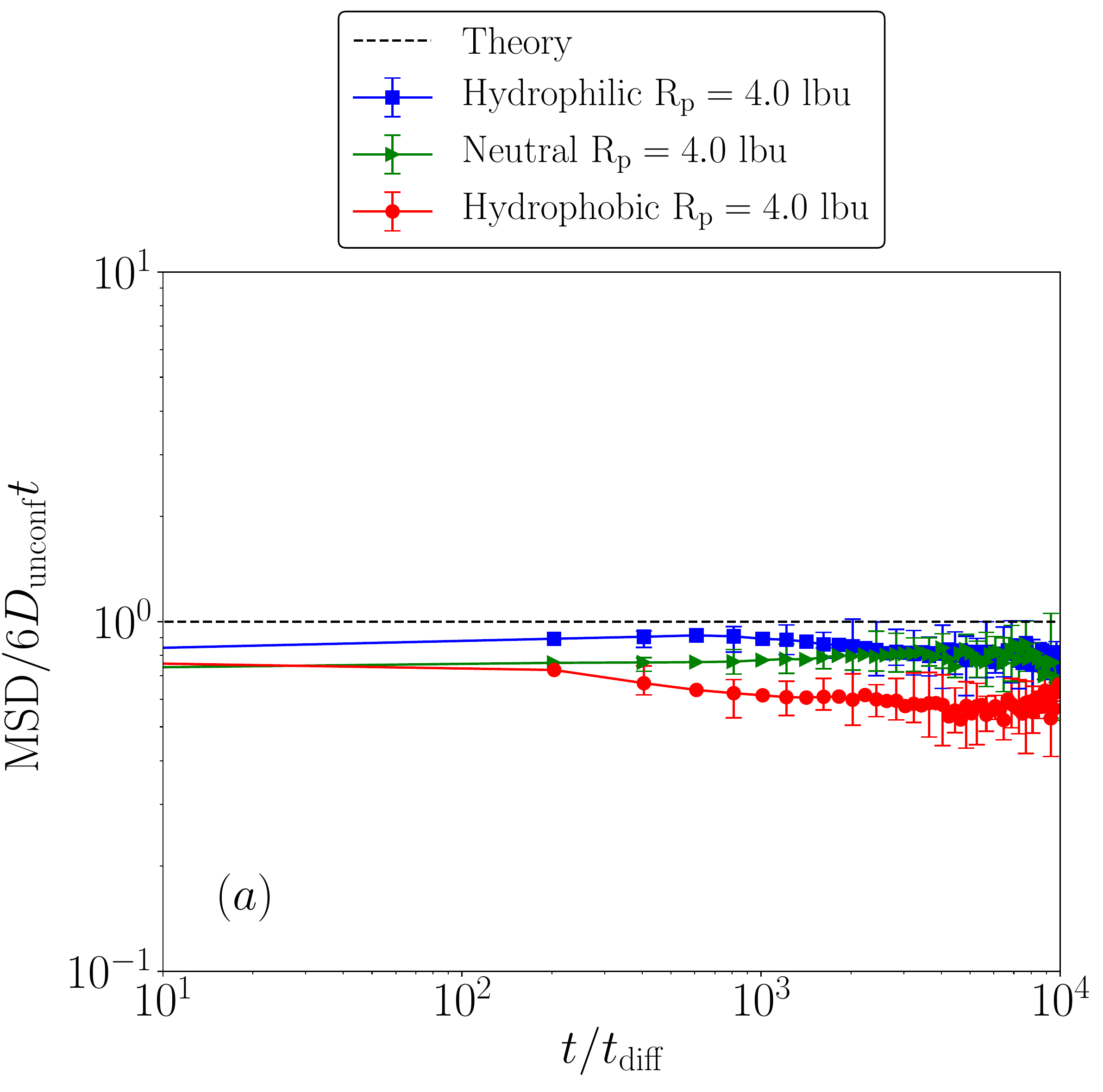}\label{fig:normalBreakPdf}}\\
\end{minipage}
\caption{The normalized Brownian particle's MSD as function of the dimensionless time $t/t_{\mathrm{diff}}$ at changing  wettabilities. Results for the MSD are normalized with the Einstein's prediction for an unconfined domain. Errorbars are the standard deviation estimated from different Brownian motion experiments.}
\label{fig:particle_brownian_unconf}
\end{figure}

We start by analyzing a case with periodic boundary conditions~\cite{ladd1993short}: the particle is initially placed in the center of the cubic box and is kicked by the fluctuating solvent.~\cref{fig:particle_msd_periodic} presents the MSD as a function of the dimensionless time $t/\tdiff$ at changing wetting parameters $w$. For $t/\tdiff<100$, we observe a neat convergence towards the expected theoretical result. However, the convergent behavior is lost when $t/\tdiff > 100$, and the MSD eventually deviates from the Einstein's relation. The reason for the mismatch is traced back to the fact that the particle is modeled on the lattice, and its movement gives rise to a ``cover-uncover" behavior on the lattice itself~\cite{ladd1993short, jansen2011bijels}. {The cover-uncover behaviour, by definition, introduces spurious momentum contributions, because when nodes are covered, some others are uncovered, and there is inevitably an adaptation dynamics that is needed for the new uncovered nodes to ``equilibrate'' with the surrounding fluid. These spurious momentum contributions should be small for large particles; unfortunately, particles need to be sufficiently small to make the Brownian diffusion simulation affordable in terms of computational time. Consequently, after some time, the system will develop a spurious advection flow in a random direction. The random advection is removed if we design a new set-up, where we place walls with a neutral wetting boundary condition at the boundaries.}
\begin{figure}[h!]
\centering
\includegraphics[width=0.7\textwidth]{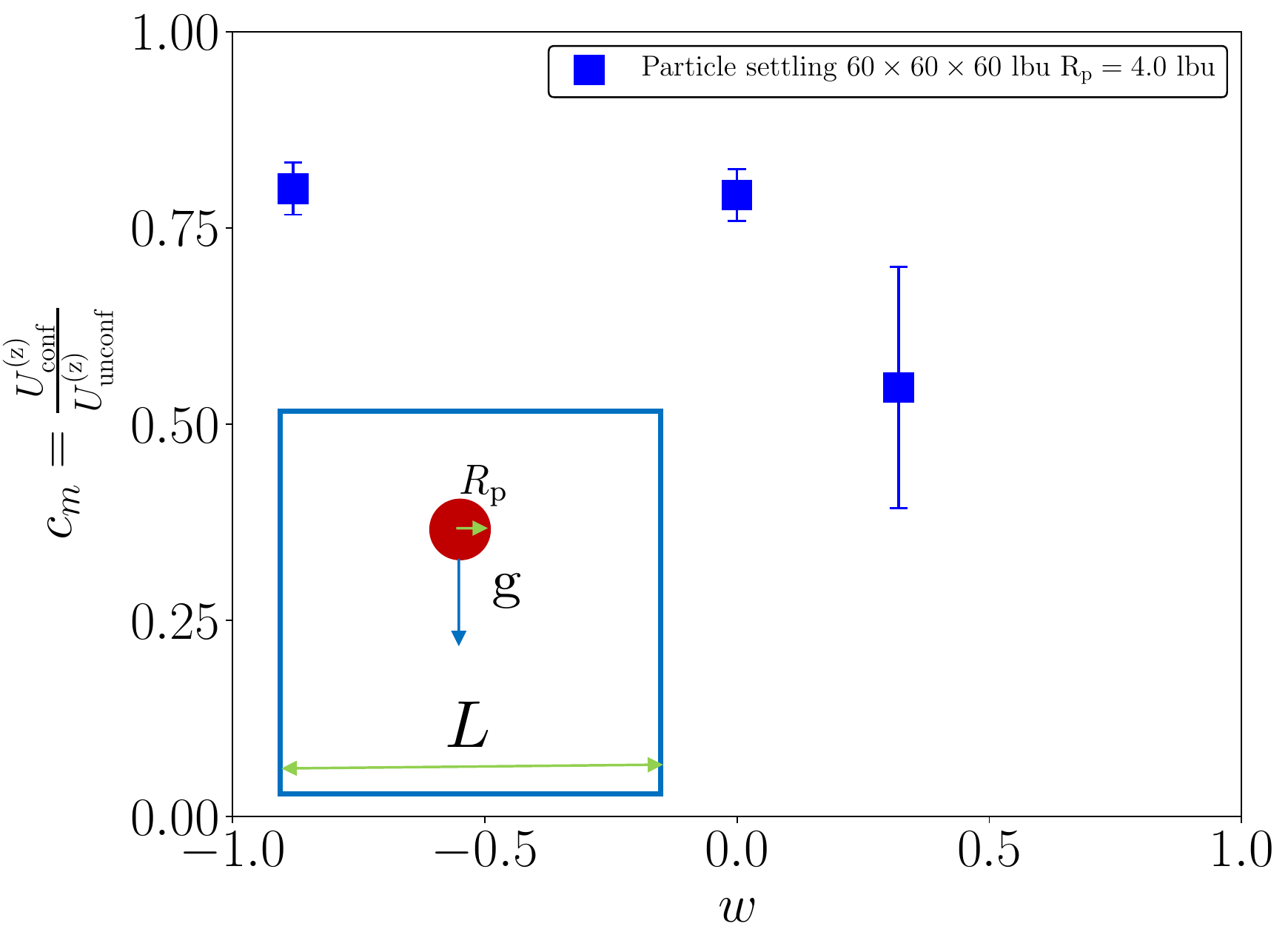}
\caption{Settling simulations: we consider particle settling in a cubic box with domain size $L^3$. The particle with radius $\rp$ is falling under the gravity $\fg$ (see inset). In the main panel we report $\cm$ (see~\cref{eq:cm}) as a function of the tunable wettability parameter $w$. We analyze particle settling simulations in three cases:  $w=-0.88$ (hydrophilic), $w=0$ (neutral), and $w=0.32$ (hydrophobic). The errorbars are the standard deviation of $\cm$ estimated from different simulations.}\label{fig:cm_measure}
\end{figure}
{Results of numerical simulations with this new set-up are reported in~\cref{fig:particle_brownian_unconf}. Results are normalized by the expected asymptotic result for an unconfined domain. We observe, however, that the normalized MSD for the three wetting conditions does not approach a unitary value for large times.}
{To delve deeper into the observed discrepancy, we notice that the particle can experience confinement effects~\cite{xue2019}.} Hence, the theoretical estimate of the diffusion coefficient should be performed by including the friction in the presence of the bounding walls. The diffusion coefficients for the confined and unconfined domains, which are $\Dconf$ and $\Do$ respectively, are defined as
\be\label{eq:diffusion_coe}
\Dconf = \frac{\kbt}{\gconf}, \hspace{.2in} \Do = \frac{\kbt}{\go}, 
\ee
where $\gconf$ is the friction of the solvent in the confined situation, and the friction $\go$ is for the unconfined case. The friction for the unconfined domain can be calculated through the Stoke's law. For the friction under confinement, the friction $\gconf$ can be calculated as $\gconf=\fp/\uconf$, where $\fp$ is the external forcing acting on the particle, and $\uconf$ is the drift velocity which can be obtained from a dedicated simulation. We thus performed particle settling simulations under confinement without thermal fluctuations~\cite{miyamura1981experimental} for hydrophilic, neutral and hydrophobic cases. The domain size is the same as the Brownian motion simulation which is $60^3$ lbu. We place the particle in the center of the domain, and apply the gravity acceleration $\fg = 10^{-6}$ lbu in the $z$ direction. After a transient time, the particle's velocity reaches a stationary state, which is defining the drift velocity. We then define $\cm$ as the ratio between the particle's drift velocity under confinement and the Stokes' prediction for an unconfined domain driven by the same volume force:
\be\label{eq:cm}
\cm=\frac{\uconf}{\uo}.
\ee
Figure~\ref{fig:cm_measure} shows the ratio $\cm$ as function of the wetting parameter $w$. Three wetting parameters, which are $w=-0.88$, $w=0$,$w=0.32$ representing hydrophilic, neutral and hydrophobic cases respectively, have been investigated. As we can see, the ratio $\cm$ is always smaller than 1, indicating that the friction under confinement is larger than the Stokes's friction. {Also, we observe that a larger errorbar appears in the hydrophobic case, which indicates that the particle is more sensitive to external fluid kicks in this situation.}
\begin{figure}[t]
\begin{minipage}{1.0\textwidth}
\centering
\subfigure{\includegraphics[width = 0.7\linewidth]{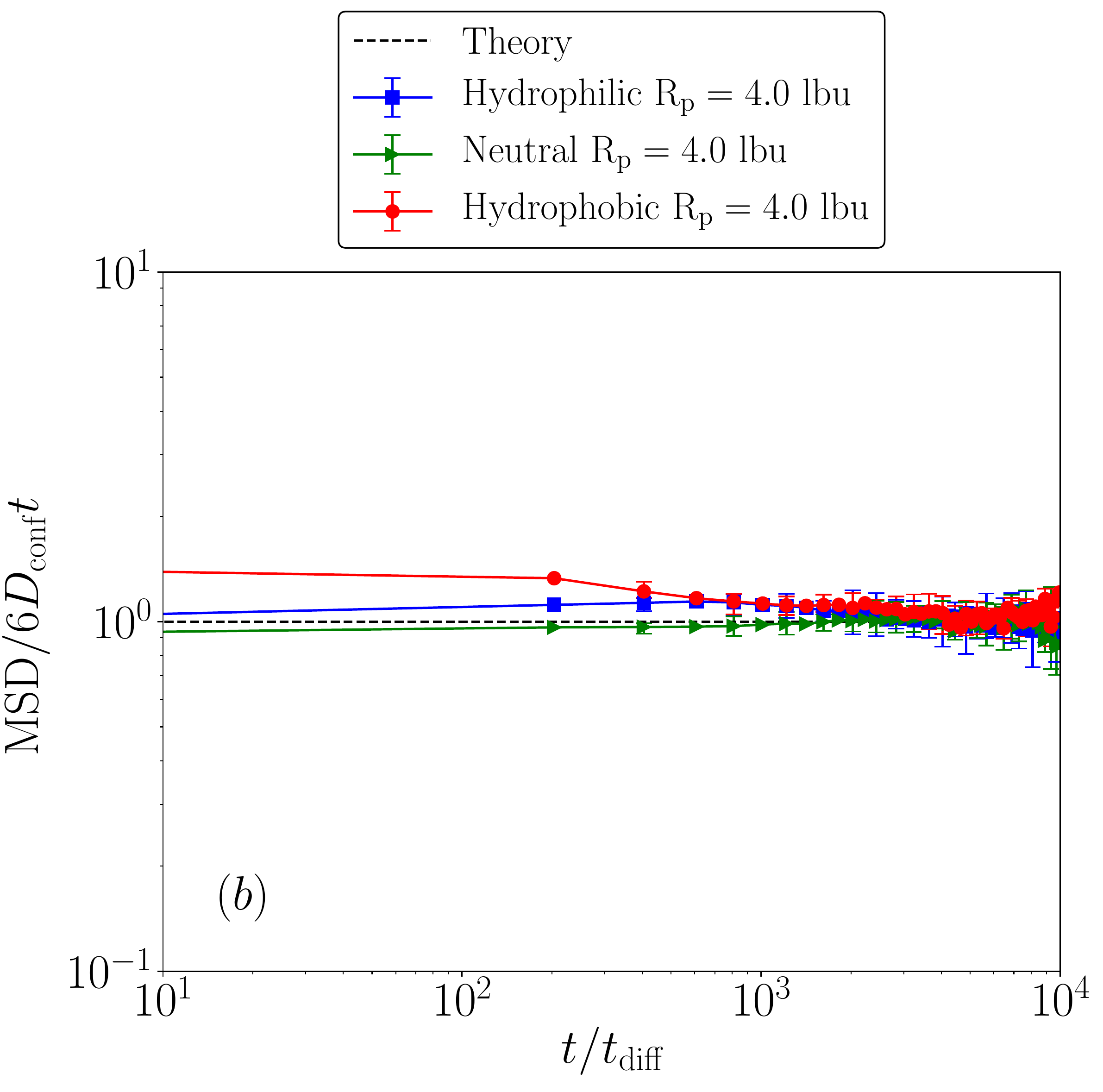}\label{fig:tbreakLaw}}\\
\end{minipage}
\caption{The normalized Brownian particle's MSD as function of the dimensionless time $t/t_{\mathrm{diff}}$ at changing  wettabilities. {The MSD normalized with the Einstein's prediction by substituting the diffusivity for unconfined domains with the diffusivity measured in confined domains. Errorbars are the standard deviation estimated from different Brownian motion experiments.}}\label{fig:particle_brownian_wall}
\end{figure}
According to~\cref{eq:diffusion_coe} and~\cref{eq:cm}, the diffusion coefficient $\Dconf$ can be obtained as:
\be\label{eq:calidiffusive}
\Dconf = \frac{\go}{\gconf} \Do=\cm\Do.
\ee
Consequently, the theoretical expectation for the diffusion coefficient in the presence of confinement is smaller than the unconfined result. {In~\cref{fig:particle_brownian_wall} we report the normalized MSD as function of $t/\tdiff$ at changing wetting conditions. At difference with respect to~\cref{fig:particle_brownian_unconf}, we now normalize the MSD with the diffusion prediction based on $\Dconf$. Now, the simulation results approach a unitary value for the three wetting conditions.}
{Summarizing, the presence of the bounding box with walls allows to remove spurious random advection; however, this introduces small -- but measurable -- confinement effects. These effects need to be taken into account for a quantitative matching with the diffusion prediction at large times.}
\section{CONCLUSIONS}\label{sec:conclusions}
In this paper, we simulated Brownian diffusion of a wetted particle with radius $\rp$ in a cubic box $L\times L \times L$ filled with the fluctuating binary mixture. The solvent is simulated with the multicomponent FLBM~\cite{Belardinelli15}, and different wettabilities of the particle have been considered based on~\cite{jansen2011bijels}. To quantitatively understand the Brownian motion of the finite-size particle, we have tested two boundary conditions on the cubic box: periodic and neutrally wetted walls. For the periodic set-up, the particle's motion experienced a random advection flow which {causes a deviation} from the Einstein's relation, due to the ``cover-uncover" behavior triggered by the particle's motion~\cite{jansen2011bijels}. The use of neutrally wetted walls removes such pathology; however, in the presence of confinement, the friction coefficient needs to be corrected to allow for a precise matching with the Einstein's relation. On a future perspective, {one could consider the present contribution a precursor study for the more complicated situation where the wetted finite size particle rests at the interface separating 2 immiscible fluids~{\cite{boniello2015brownian}}. Also, further investigations on the dynamics of Brownian particles in shear flow to study the Taylor dispersion could be interesting.} Moreover, we would like to mention that the particle is still not very well {resolved}. Hence, it could be a challenging computational task to perform numerical simulations with larger resolutions to tackle these kind of problems. 
\section{ACKNOWLEDGEMENTS}
The authors would like to kindly acknowledge funding from the European Union's Horizon 2020 research and innovation programme under the Marie Sk\l{}odowska-Curie grant agreement No. 642069 (European Joint Doctorate Programme ``HPC-LEAP").


\bibliography{prex}

\end{document}